\documentclass[10pt,aps,twocolumn,amsmath,amssymb]{revtex4-1}
\usepackage{geometry} 

\usepackage{graphicx}
\usepackage{amssymb,amsmath}
\usepackage{epstopdf}
\usepackage{comment}
\usepackage[dvipdfmx]{color}
\def\eq#1{\begin{equation}#1\end{equation}}

\def\HH{H_{\mathrm{tdPU}}}

\def\DDt#1{\frac{\mathrm{d}#1}{\mathrm{d}t}}

\def\del#1#2{\frac{\partial #1}{\partial #2}}
\DeclareSymbolFont{lettersA}{U}{txmia}{m}{it}
\DeclareMathSymbol{\FFF}{\mathord}{lettersA}{'206}
\DeclareMathSymbol{\NNN}{\mathord}{lettersA}{'216} 
\DeclareMathSymbol{\RRR}{\mathord}{lettersA}{'222} 
\DeclareMathSymbol{\ZZZ}{\mathord}{lettersA}{'232} 
\DeclareMathSymbol{\QQQ}{\mathord}{lettersA}{'221} 
\DeclareMathSymbol{\CCC}{\mathord}{lettersA}{'203} 

\begin{document}

\title{Time dependent Pais-Uhlenbeck oscillator and its decomposition
}

\author{Hirosuke Kuwabara}
\email[]{kuwabara-hirosuke@ed.tmu.ac.jp}
\affiliation{Department of Physics, Graduate School of Science and Engineering, Tokyo Metropolitan University, Hachioji-shi, Tokyo 192-0397, Japan}
\author{Tsukasa Yumibayashi}
\email[]{yumibayashi-tsukasa@ed.tmu.ac.jp}
\affiliation{Department of Physics, Graduate School of Science and Engineering, Tokyo Metropolitan University, Hachioji-shi, Tokyo 192-0397, Japan}
\author{Hiromitsu Harada}
\email[]{harada-hiromitsu@ed.tmu.ac.jp}
\affiliation{Department of Physics, Graduate School of Science and Engineering, Tokyo Metropolitan University, Hachioji-shi, Tokyo 192-0397, Japan}

\begin{abstract}
The Pais-Uhlenbeck(PU) oscillator is the simplest model with higher time derivatives. Its properties were studied for a long time.
In this paper, we extend the 4th order free PU oscillator to a more non-trivial case, dubbed the 4th order time dependent PU oscillator, which has time dependent frequencies. 
We show that this model cannot be decomposed into two harmonic oscillators in contrast to the original PU oscillator. An interaction is added by the coordinate transformation of Smilga. 
\end{abstract}

\maketitle

\section{Introduction}
It is worth investigating higher derivative theories. A higher derivative theory usually violates unitarity, and the Pais-Uhlenbeck(PU) oscillator\cite{PU} is the easiest model for studies what the higher derivatives cause.
In higher derivative theories, their Hamiltonian needed for quantization and examining stability can be obtained by Ostrogradski's method\cite{O,KMY}. The Hamiltonian of the PU oscillator looks little complicated, but it can be rewritten to the very simple form of several independent harmonic oscillators. 
For example, as regards the 4th order PU oscillator, whose equation of motion(EOM) has up to the 4th order time derivative, its Hamiltonian $H_{\mathrm{PU}}$ can be separated into two independent harmonic oscillator Hamiltonians, $H_{\mathrm{PU}} = H_1 - H_2$, by Smilga's canonical transformation\cite{S}.
In the last form, quantization is quite simple, just by quantizing each of harmonic oscillators. 
It is obvious that the negative sign before $H_2$ means that the energy is not bounded from below, so that
this model may be unstable (once interactions are added)\cite{N,P}. The negative energy generically comes out in higher derivative theories\cite{W}, however, for the PU oscillator model no instability occurs because it is a free model\cite{HH,W,S,B,BM}. Given some interactions, maintaining separability is known to play the key role for the stability of the interacting  PU model. It is significant to consider other interacting higher derivative PU models, and examine whether the separability is conserved or not.
In this paper, we extend the 4th order PU oscillator to the time dependent 4th order PU(tdPU) oscillator, whose frequencies depend on time, to see whether it has the separability or not. Our model may be considered as a PU
oscillator under the influence of some external forces, so that the energy is not conserved.
In this situation, when one naively uses Smilga's transformation, a problem arises. The tdPU Hamiltonian obtained by Ostrogradski's method and the Hamiltonian transformed by Smilga are in one to one correspondence with each other, however, their equations of motion are not. This means that Smilga's transformation is not canonical in our model, so that it is necessary to add some correction terms to the naively transformed Hamiltonian.
Here we find those correction terms from the two points of view: (i) a comparison of the differential equations, and (ii) the generating function method. We show that the separability of our model is deformed because the corrections become the interaction terms of two time dependent harmonic oscillators.

\section{Previous Results}
\subsection{4th order PU oscillator}
In \cite{PU, S}, the 4th order PU oscillator is defined as
\eq{
L_{\mathrm{PU}}(x,\dot{x},\ddot{x}) = \frac{1}{2} \ddot{x}^2 - \frac{1}{2} \Omega_2 \dot{x}^2 + \frac{1}{2} \Omega_1 x^2,
\label{eq:pu2}
}
where $\Omega_1:= \omega_1^{\ 2} \omega_2^{\ 2}, \Omega_2:= \omega_1^{\ 2}+ \omega_2^{\ 2}$.
Let us introduce the PU Hamiltonian by applying Ostrogradski's method\cite{O} to (\ref{eq:pu2}).
The PU Hamiltonian $H_{\mathrm{PU}}$ is 
\begin{eqnarray}
\label{eq:puh}
H_{\mathrm{PU}}(q_1, q_2, p_1, p_2 ) &=& p_1 \dot{q_1} + p_2 \dot{q_2} - L_{\mathrm{PU}} \nonumber \\
                          &=& p_1 q_2 + \frac{1}{2} p_2 ^{\ 2}  \nonumber \\
                              && + \frac{1}{2} \Omega_2 q_2^{\ 2} -\frac{1}{2} \Omega_1 q_1^{\ 2},
\end{eqnarray}
where $q_i, p_i \ (i=1,2)$ are canonical coordinates,
\begin{equation}
\label{eq:ltrans}
\begin{cases}
q_1 = x,\\
p_1 = \displaystyle \del{L_{\mathrm{PU}}}{\dot{x}} - \DDt{} \del{L_{\mathrm{PU}}}{\ddot{x}} \\
\ \quad= - \Omega_2 \dot{x} - \dddot{x}, \\
q_2 = \dot{x}, \\
p_2 = \displaystyle \del{L_{\mathrm{PU}}}{\ddot{x}} = \ddot{x}.
\end{cases}
\end{equation}

\subsection{Decomposition of the PU Hamiltonian by Smilga's transformation}
In \cite{S}, Smilga gave the transformation that separates $H_{\mathrm{PU}}$ as follows:
\begin{equation}
\label{eq:strans}
\begin{cases}
q_1 = \displaystyle \frac{\gamma}{\omega_1} (\omega_1 Q_2 - P_1),\\
p_1 = \displaystyle \omega_1\gamma (\omega_1 P_2 - \omega_2^{\ 2} Q_1), \\
q_2 = \displaystyle \gamma(\omega_1 Q_1 - P_2), \\
p_2 = \displaystyle \gamma (\omega_1 P_1 - \omega_2^{\ 2} Q_2).
\end{cases}
\end{equation}
where $\gamma:= 1/\sqrt{\omega_1^{\ 2}- \omega_2^{\ 2}}$. In this paper, we call it Smilga transformation.

Here we assume $\omega_1 > \omega_2$ for simplicity, and this transformation is a canonical transformation. According to (\ref{eq:strans}), $H_{\mathrm{PU}}$ can be rewritten to
\begin{eqnarray}
\label{eq:spu}
H_{\mathrm{PU}}(Q_1, Q_2, P_1, P_2) &=& H_1(Q_1, P_1) \nonumber \\
&& -H_2(Q_2, P_2),
\end{eqnarray}
where $H_i(Q_i, P_i):= \frac{1}{2} P_i^{\ 2} + \frac{1}{2} \omega_i^{\ 2} Q_i^{\ 2}$.

Thus we see that the PU oscillator can be separated into two harmonic oscillators. Smilga used this property to quantize the PU oscillator and and investigate its quantum stability.

\section{Our results}
\subsection{Time dependent PU oscillator}
Our Lagrangian of the time dependent PU(tdPU) oscillator is 
\begin{eqnarray}
L_{\mathrm{tdPU}}(x,\dot{x},\ddot{x}) &=& \frac{1}{2} \ddot{x}^2 - \frac{1}{2}\Omega_2(t) \dot{x}^2 \nonumber \\
&& + \frac{1}{2} \Omega_1(t) x^2.
\end{eqnarray}
where $\Omega_1(t):= \omega_1^{\ 2}(t) \omega_2^{\ 2}(t), \Omega_2(t):= \omega_1^{\ 2}(t)+ \omega_2^{\ 2}(t)$.
This Lagrangian can be obtained by replacing $\omega_i \to \omega_i (t) (i=1,2)$ in (\ref{eq:pu2}).
Its EOM is
\begin{eqnarray}
\ddddot{x} +\Omega_2(t) \ddot{x} +\dot{\Omega}_2(t) \dot{x} +\Omega_1(t) x = 0.
\label{eq:Leom}
\end{eqnarray}

The Hamiltonian and its EOM are 
\begin{eqnarray}
\label{eq:tdpu}
\HH(q_1, q_2, p_1, p_2) &=& p_1 \dot{q_1} + p_2 \dot{q_2} - L_{\mathrm{tdPU}} \nonumber \\
                          &=& p_1 q_2 + \frac{1}{2} p_2 ^{\ 2} \nonumber \\
                          && + \frac{1}{2} \Omega_2(t) q_2^{\ 2} \nonumber \\
                          && - \frac{1}{2} \Omega_1(t) q_1^{\ 2},
\end{eqnarray}
\begin{eqnarray}
\label{eq:Heom1}
\DDt{} \left(
    \begin{array}{c}
      q_1 \\
      q_2 \\
      p_1 \\
      p_2
    \end{array}
  \right) &=& 
A
\left(
    \begin{array}{c}
      q_1 \\
      q_2 \\
      p_1 \\
      p_2
    \end{array}
  \right),
\end{eqnarray}
where
\begin{equation}
A :=
 \left(
    \begin{array}{cccc}
      0 & 1 & 0 & 0 \\
      0 & 0 & 0 & 1 \\
      \Omega_1(t) & 0 & 0 & 0 \\
      0 & -\Omega_2(t) & -1 & 0
    \end{array}
  \right).
\label{eq:A}
\end{equation}
Here $p_i,q_i \ (i=1,2)$ are defined by (\ref{eq:ltrans}) with time dependent frequencies, (\ref{eq:Leom}) and (\ref{eq:Heom1}) are equivalent in our case.

By using (\ref{eq:strans}) with time dependent frequencies, (\ref{eq:tdpu}) can be written as
\begin{eqnarray}
\label{eq:HtdPU}
\hspace{-0.5cm}
&&\HH(Q_1, Q_2, P_1, P_2) \nonumber \\
&&= (H_\mathrm{td})_1(Q_1, P_1)  - (H_\mathrm{td})_2(Q_2, P_2),
\end{eqnarray}
where $(H_\mathrm{td})_i(Q_i, P_i)=\frac{1}{2} P_i^{\ 2} + \frac{1}{2} \omega_i^{\ 2}(t) Q_i^{\ 2}$.
Its EOM is
\begin{eqnarray}
\label{eq:Heom2}
\DDt{} \left(
    \begin{array}{c}
      Q_1 \\
      Q_2 \\
      P_1 \\
      P_2
    \end{array}
  \right) &=& 
B
\left(
    \begin{array}{c}
      Q_1 \\
      Q_2 \\
      P_1 \\
      P_2
    \end{array}
  \right). \\
\nonumber
\end{eqnarray}
where
\begin{equation}
B := 
 \left(
    \begin{array}{cccc}
      0 & 0 & 1 & 0 \\
      0 & 0 & 0 & -1 \\
      -\omega^{\ 2}_1(t) & 0 & 0 & 0 \\
      0 & \omega^{\ 2}_2(t) & 0 & 0
    \end{array}
  \right),
\end{equation}
It is straightforward to check that (\ref{eq:tdpu}) is equal to (\ref{eq:HtdPU}) under the Smilga transformation.
However, the time dependent Smilga transformation is not a canonical transformation. In other words, it doesn't reproduce the EOM (\ref{eq:Leom}) or, equivalently, (\ref{eq:Heom1}) does not follow from (\ref{eq:Heom2}).
In the next two Sections, we show that some correction terms are needed to correct the situation, and they become interaction terms of the harmonic oscillators.

\subsection{Comparison of EOMs}
Here we find the correction term by a comparison of the EOMs (\ref{eq:Heom1}) and (\ref{eq:Heom2}).
Let us introduce some notation first. 
\begin{eqnarray}
X &:=& (q_1, q_2, p_1, p_2)^{T}, \\
Y &:=& (Q_1, Q_2, P_1, P_2)^{T},
\end{eqnarray}
Here $^{T}$ denotes transpose of a matrix.
$Y, X$ and $A, B$ are related by 
\begin{eqnarray}
\label{eq:rel}
Y &=& M X, \label{eq:rel1}\\ 
MA &=& BM \label{eq:rel2},
\end{eqnarray}
where $M$ is the (inverse) coefficient matrix of the Smilga transformation (\ref{eq:strans}).
With these notation, the EOM (\ref{eq:Heom1}) and (\ref{eq:Heom2}) can be written as
\begin{eqnarray}
\DDt{} X &=& A X, \label{eq:eom1}\\
\DDt{} Y &=& B Y + \alpha. \label{eq:eom2}
\end{eqnarray}
Here $\alpha$ represents a correction to (\ref{eq:Heom2}).
Differentiating the first relation (\ref{eq:rel1}) with respect to $t$ and substituting (\ref{eq:eom2}) to the left hand side, we find
\begin{eqnarray*}
\DDt{} Y &=& M \DDt{} X + \dot{M} X = M A X + \dot{M} X \\
&=& BY + \alpha = BMX + \alpha, \\
\therefore \alpha &=& \dot{M} X = \dot{M} M^{-1} Y. \\
\end{eqnarray*}
Therefore, $\alpha \neq 0$, unless the model is time independent. More explicitly, $\alpha$ is
\begin{eqnarray*}
\alpha &=&  \gamma^2(t)
 \left(
    \begin{array}{cccc}
     C_1 & 0 & 0 & E_1 \\
      0 & C_2 & E_1 & 0 \\
      0 & E_2 & C_3 & 0 \\
     E_2 & 0 & 0 & C_4
    \end{array}
  \right) Y,
\end{eqnarray*}
where 
\begin{eqnarray*}
C_1 &:=&  \frac{\dot{\gamma}(t)}{\gamma^3(t)}+ \frac{\dot{\omega}_1(t)}{\omega_1(t)} \Omega_2(t), \\
C_2 &:=& \frac{\dot{\gamma}(t)}{\gamma^3(t)} + 2 \omega_1(t) \dot{\omega}_1(t),\\
C_3 &:=& \frac{\dot{\gamma}(t)}{\gamma^3(t)} + \frac{\dot{\omega}_1(t)}{\omega_1(t) \gamma^2(t)} - 2  \omega_2(t) \dot{\omega}_2(t),\\
C_4 &:=& \frac{\dot{\gamma}(t)}{\gamma^3(t)} - 2 \omega_2(t) \dot{\omega}_2(t) ,\\
E_1 &:=& - 2 \dot{\omega}_1(t),\ E_2:=2 \omega_1(t) \omega_2(t) \dot{\omega}_2(t), \\
\gamma(t) &:=& 1/\sqrt{\omega^{\ 2}_1(t) - \omega^{\ 2}_2(t)}.
\end{eqnarray*}

\newpage
\subsection{Generating function method}
In this part, we show that the correction term can be also obtained by using a canonical Hamiltonian $\HH'$.
The latter can be found by the generating function method, via solving the equations
\begin{eqnarray*}
&& p_i = \del{W(q,Q,t)}{q_i},\ \ P_i = - \del{W(q,Q,t)}{Q_i}, \\ 
&& \HH(q,p)- \HH'(Q,P) = - \del{W(q,Q,t)}{t},
\end{eqnarray*}
where $q:=(q_1, q_2), p:=(p_1, p_2), Q:=(Q_1, Q_2), P:=(P_1, P_2)$ and $W(q,Q,t)$ is the generating function of a canonical transformation. Then we find
\begin{eqnarray*}
W(q,Q,t) &=& - \omega^{\ 2}_1(t) q_1 q_2 + \frac{\omega_1(t)}{\gamma(t)} Q_1 q_1 \nonumber \\ 
             &+& \frac{1}{\gamma(t)} Q_2 q_2 - \omega_1(t) Q_1 Q_2 + g(t).
\end{eqnarray*}
Here $g(t)$ is an arbitrary time dependent function.
Using the above we find that the $\HH'(Q,P)$ is given by
\begin{eqnarray}
 \HH'(Q,P) &=& H_{\mathrm{tdPU}}(Q,P) +  \del{W}{t} \nonumber \\
                &=& H_{\mathrm{tdPU}}(Q,P) +H_{\mathrm{int}} \label{eq:HtdPU2},
\end{eqnarray}
where 
\begin{eqnarray}
\hspace{-0.5cm} H_{\mathrm{int}} &:=& - 2\gamma^2(t) \omega_1(t) \omega_2(t) \dot{\omega}_2(t) Q_1 Q_2 \nonumber \\
&& - 2\gamma^2(t) \dot{\omega}_1(t)P_1 P_2 \nonumber \\
&& + \frac{\gamma^2(t) \omega_2(t)}{\omega_1(t)} (\omega_1(t) \dot{\omega}_2(t) + \dot{\omega}_1(t) \omega_2(t)) P_1 Q_1 \nonumber \\
&& + \gamma^2(t) (\omega_1(t) \dot{\omega}_1(t) + \dot{\omega}_2(t) \omega_2(t)) P_2 Q_2  \nonumber \\
&& + \dot{g}(t)
\end{eqnarray}
Its EOM is 
\begin{eqnarray*}
\DDt{} Y &=& BY 
            + \gamma^2(t)
 \left(
    \begin{array}{cccc}
     F_1 & 0 & 0 & E_1 \\
      0 & F_2&E_1 & 0 \\
      0 & E_2 &F_3& 0 \\
     E_2 & 0 & 0 & F_4
    \end{array}
  \right) Y,
\end{eqnarray*}
where 
\begin{eqnarray*}
F_1 &=& \frac{\omega_2(t)}{\omega_1(t)}\ (\omega_1(t) \dot{\omega}_1(t) + \omega_2(t) \dot{\omega}_1(t)), \\
F_2 &=& \omega_1(t) \dot{\omega}_1(t) + \omega_2(t) \dot{\omega}_2(t),\\
F_3 &=&- \frac{\omega_2(t)}{\omega_1(t)}\ (\omega_1(t) \dot{\omega}_2(t) + \omega_2(t) \dot{\omega}_1(t)), \\
F_4 &=& - (\omega_1(t) \dot{\omega}_1(t) + \omega_2(t)  \dot{\omega}_2(t)).
\end{eqnarray*}
It is easily confirmed that $C_i = F_i,\ (i=1,2,3,4)$. This means that the new terms in (\ref{eq:HtdPU2}) reproduce $\alpha$. Hence, the interacting Hamiltonian (\ref{eq:HtdPU2}) is canonical and the correction $\alpha$ of the EOM (\ref{eq:Heom2}) is equivalent to the interaction terms in the canonical Hamiltonian (\ref{eq:HtdPU2}).


\section{Conclusion}
In this paper, we extended the free 4th order PU oscillator to the non-trivial case, the 4th order tdPU oscillator, whose frequencies depend on time. We found that our model cannot be written down in the form of two harmonic oscillators. This is because the Smilga transformation is not canonical in our extended model, so that  it was necessary to add some correction terms to make it canonical. We obtained those corrections from the two points of view: (i) a comparison of the differential equations, and (ii) the generating function method.
We showed that the correction terms in the tdPU EOM can be written as interaction terms in the canonical tdPU Hamiltonian.

\section*{Acknowledgements}
The authors thank S. V. Ketov, A. V. Smilga and S. L. Lyakhovich for useful discussions and correspondence.

\newpage


\end{document}